\documentclass[english,aps,prl,twocolumn,superscriptaddress,showpacs, amsmath]{revtex4-1}

\usepackage[T1]{fontenc}
\usepackage[latin9]{inputenc}
\usepackage{pdfpages}
\usepackage{color}
\usepackage{units}
\usepackage{amssymb}
\usepackage{graphicx}
\usepackage{esint}
\usepackage{bm}
\usepackage{natbib}
\usepackage{amsmath}

\usepackage{ulem}
\setcitestyle{journalcolor= blue}

\usepackage{babel}

\makeatletter
\AtBeginDocument{\let\LS@rot\@undefined}
\makeatother

\begin{document}
\title{Inter-Landau-level Andreev Reflection at the Dirac Point in a Graphene Quantum Hall
State Coupled to a NbSe$_2$ Superconductor}

\author{Manas Ranjan Sahu}
\affiliation{Department of Physics, Indian Institute of Science, Bangalore 560012, India}
\author{Xin Liu}
\affiliation{School of Physics, Huazhong University of Science and Technology, Wuhan 430074, China}
\author{Arup Kumar Paul}
\affiliation{Department of Physics, Indian Institute of Science, Bangalore 560012, India}
\author{Sourin Das}
\affiliation{Indian Institute of Science Education and Research, Kolkata, Mohanpur 741246, India}
\author{Pratap Raychaudhuri}
\affiliation{Tata Institute of Fundamental Research, Homi Bhabha Road, Colaba, Mumbai 400 005, India}
\author{J. K. Jain}
\affiliation{Department of Physics, The Pennsylvania State University, University Park, Pennsylvania 16802, USA}
\author{Anindya Das}
\email{anindya@iisc.ac.in}
\affiliation{Department of Physics, Indian Institute of Science, Bangalore 560012, India}

\begin{abstract}
Superconductivity and quantum Hall effect are distinct states of matter occurring in apparently incompatible physical conditions. Recent theoretical developments suggest that the coupling of quantum Hall effect with a superconductor can provide a fertile ground for realizing exotic topological excitations such as non-abelian Majorana fermions or Fibonacci particles. As a step toward that goal, we report  observation of Andreev reflection at the junction of a quantum Hall edge state in a single layer graphene and a quasi-two dimensional niobium diselenide (NbSe$_2$) superconductor.  Our principal finding is the observation of an anomalous finite-temperature conductance peak located precisely at the Dirac point, providing a definitive evidence for inter-Landau level Andreev reflection in a quantum Hall system. Our observations are well supported by detailed numerical simulations, which offer additional insight into the role of the edge states in Andreev physics. This study paves the way for investigating analogous Andreev reflection in a fractional quantum Hall system coupled to a superconductor to realize exotic quasiparticles.
\end{abstract}

\maketitle

\begin{figure*}[ht!]
\includegraphics[width=1\textwidth]{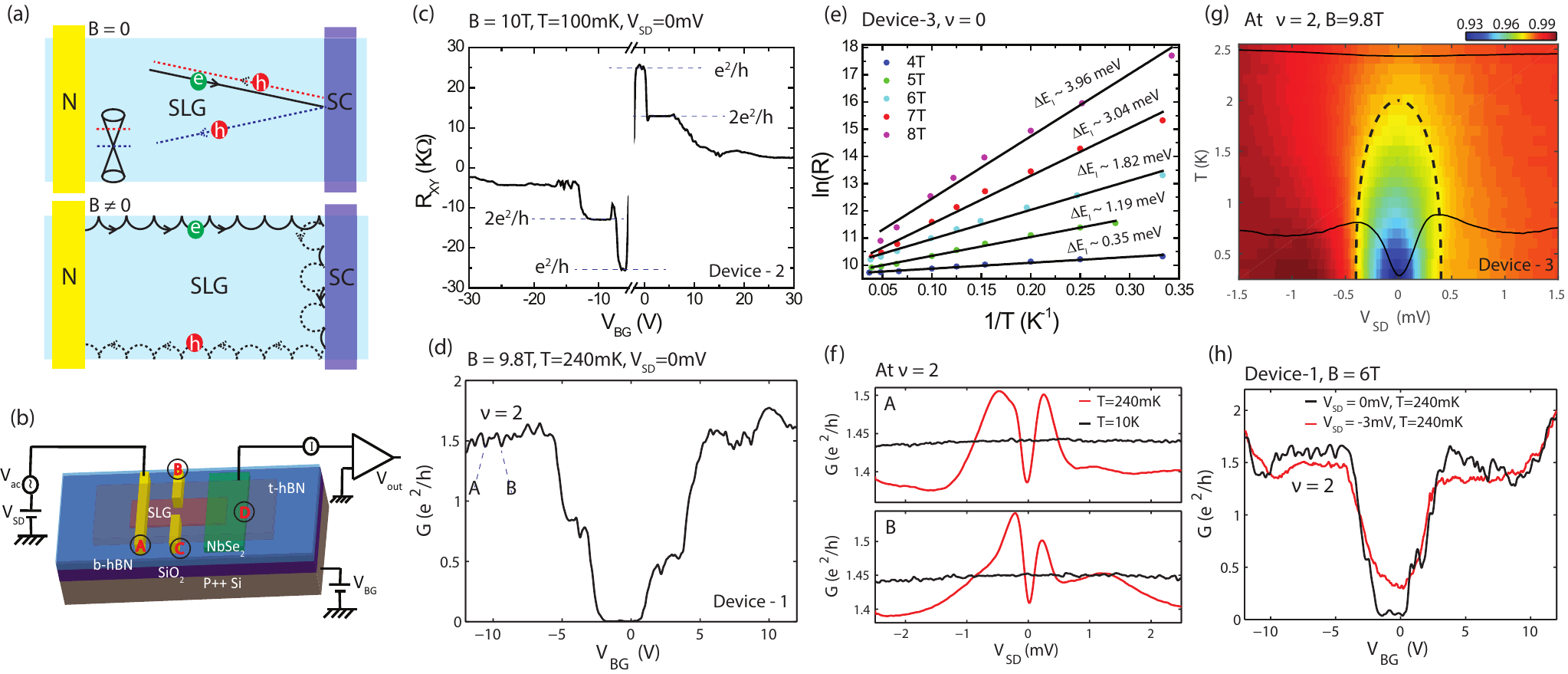}
 \caption{(Color Online) (\textbf{a}) (top) AR in graphene at B = 0. The red (blue) dashed line shows retro (specular) AR. (bottom) Classical picture of AR at the interface of QH edge state and superconductor based on skipping orbit. The electron and hole orbits have the same chirality for intra-band process. (\textbf{b}) Schematic of the experimental measurement setup of hBN protected graphene devices. For R$_{\rm xy}$ measurement current is injected between A and D, voltage is measured between B and C. For the two probe conductance measurement of the SLG-NbSe$_2$ junction voltage is applied at A, and current is measured at D. (\textbf{c}) R$_{\rm xy}$ of device 2 at B = 10T showing symmetry broken QH plateaus. (\textbf{d}) Two-terminal gate response of device 1 between Au-SLG-NbSe$_2$ at B = 9.8T and V$_{\rm SD}$ = 0mV. (\textbf{e}) Activation plot for device 3 at the Dirac point for different magnetic fields; the corresponding insulating gaps are shown on the figure. We note that the resistance changes by up to three orders of magnitude over the range of the fits. (\textbf{f}) $dI/dV$ as a function of V$_{\rm SD}$ measured in device 1 at B = 9.8T on the $\nu$ = 2 LL at the positions A and B marked in fig(d); BCS peaks are present at 240 mK (red) but not at 10K (black). (\textbf{g}) 2D colormap of normalized $dI/dV$ versus V$_{\rm SD}$ as a function of temperature at B = 9.8T for device 3. Superconductivity vanishes at around 2K. The black dashed line is the theoretical temperature dependence of BCS gap. The cut lines are shown at 240mK and 2.5K. (\textbf{h}) The gate responses of device 1 for 6T at V$_{\rm SD}=0$ (black) and for $|$eV$_{\rm SD}|>\Delta$ (red). The former has enhanced conductance.}
 \label{fig1}
\end{figure*}

Proximity effect through Andreev reflection (AR) is the primary ingredient for engineering a topological superconductor, which is expected to be a breeding ground for new types of topological excitations\cite{fu2008superconducting,lutchyn2010majorana,oreg2010helical,mourik2012signatures,das2012zero,mong2014universal,alicea2016topological,clarke2014exotic}. Discovery of graphene in the last decade\cite{novoselov2004electric}, aided by developments in improving device quality by encapsulating with hexagonal Boron Nitride\cite{wang2013one,kim2009large} (hBN), provides one of the best opportunities to extend the study of AR for Dirac electrons in proximity to superconductor\cite{beenakker2006specular,beenakker2008colloquium,calado2015ballistic,allen2016spatially,efetov2015specular,sahu2016andreev,han2014collapse,soori2018enhanced}.
In these systems an incident electron from the single layer graphene (SLG) with a finite excitation energy combines with another electron below the Fermi energy ($E_F$) to form a Cooper pair at the junction (Fig. 1a-top). 
The AR and its transition from retro to non-retro reflection has been observed 
 \cite{sahu2016andreev}. More interestingly, when $E_F$ is aligned with the Dirac point, AR requires an inter-band process and is predicted to be specular (Fig. 1a-top), as observed recently in bilayer graphene\cite{efetov2015specular}.

Exotic physics is predicted to arise from the coupling between a superconductor and a topological quantum Hall (QH) state. In particular, this system has been proposed as a novel route for creating a variety of non-abelian anyons,
which have been hailed as possible building blocks for future topological quantum computation\cite{bishara2007non,nayak2008non,mong2014universal}. 
The physics of AR is predicted to alter dramatically in the QH regime \cite{hoppe2000andreev,fisher1994cooper,PhysRevB.72.054518}, where electron transport occurs primarily through the chiral edge states, which themselves are topologically robust manifestations of the Landau Levels (LLs) in the interior of the sample. On the QH plateau, an incident chiral electron is expected to bounce back as an Andreev-reflected chiral hole propagating in the same direction as the incoming electron (Fig. 1a - bottom)\cite{akhmerov2007detection}, due to the sign reversals of both the charge and the mass. 
A difficulty in experimentally investigating this physics is the fact that high magnetic fields required for the QH effect are inimical to superconductivity. Important progress has recently been made in this direction. Supercurrent and Josephson coupling in QH regime at SLG-superconductor interface have been demonstrated at relatively low magnetic field ($\sim$ 2T)\cite{amet2016supercurrent,shalom2016quantum,doi:10.1021/nl204415s}. At high magnetic fields ($\sim$ 10T) the superconducting correlations in QH edge has been realized recently\cite{lee2017inducing}.

In this work, we show that a coexistence of, and a coupling between, a QH system and a superconductor can be realized and studied in a system of SLG coupled to a NbSe$_2$ superconductor. Our results reveal that at high magnetic fields, when the breaking of the spin and valley symmetries generally fully splits the zeroth Landau level\cite{abanin2006spin,tikhonov2016emergence,kharitnov2012edge}, AR manifests most strikingly through an anomalous conductance peak located precisely at the Dirac point (DP). We attribute this peak to inter-Landau level AR, and confirm its physical origin by detailed theoretical simulations. 

Our devices consist of an SLG partially covered with a thin film of NbSe$_2$ (Fig.~1b). Details of the fabrication and measurement schemes are given in the Supplemental Material (SM)\cite{supplement} Sec. SI1. We show results from three devices as a function of the back-gate voltage ($V_{\rm BG}$), the source-drain bias voltage ($V_{\rm SD}$), the temperature ($T$) and the magnetic field (B). The highest mobility of 60,000 cm$^2$/V.sec was obtained in device-3, where the carrier inhomogeneity ($\delta$n) due to charge puddles was $\sim$ (3-5) $\times$ 10$^9$ cm$^{-2}$  which corresponds to Fermi energy broadening ($\delta$E$_F$) of $\sim$ 6-8meV
\cite{Xue2011}. The characterization of several devices is shown in SM Sec. SI1\cite{supplement}. Fig. 1c presents the Hall resistance, R$_{\rm xy}$, of device 2 at $B = 10$T, where the plateaus at 2e$^2$/h and 1e$^2$/h are clearly visible. From the B dependence of Shubnikov de Haas oscillations\cite{hong2009quantum,young2012spin} the LL broadening of $\Gamma$ $\sim$ 4.5 meV was obtained (SM Sec. SI3\cite{supplement}).  
 The two-probe conductance (G) measured between SLG - superconductor contact at 9.8T is shown in Fig. 1d (device 1). The value of conductance on the plateaus is lower than the ideal value due to the contact resistance of $\sim$ 1.5 kilo-ohms at the SLG - NbSe$_{2}$ junction. 
 In addition to different broken symmetries, an insulating state, i.e. a $\nu$ = 0 plateau, is observed at the DP
as previously reported in the literature \cite{zhang2006landau,jiang2007quantum,checkelsky2008zero,du2009fractional,bolotin2009observation}. 
Using thermally activated carrier transport model  
we have determined the insulating gap of the $\nu$ = 0 plateau (SM Sec. SI5\cite{supplement}). Previous studies\cite{bolotin2009observation,du2009fractional} have reported that the value of insulating gap of $\nu$ = 0 plateau depends on $\Gamma$, and the measured activation gap is nothing but the mobility gap, $\Delta E_I$ = $\Delta E_{LL}$ - $\Gamma$\cite{giesbers2007quantum,young2012spin}.  At 10T, $\Delta E_I \sim$ 5 meV was measured for device 3 (SM sec. SI5\cite{supplement}), and activation plots at several B are shown in Fig. 1e. The details of the activation plots of device 1 and device 2 are shown in SM Sec. SI5\cite{supplement}.

\begin{figure}[ht!]
 \includegraphics[width=0.5\textwidth]{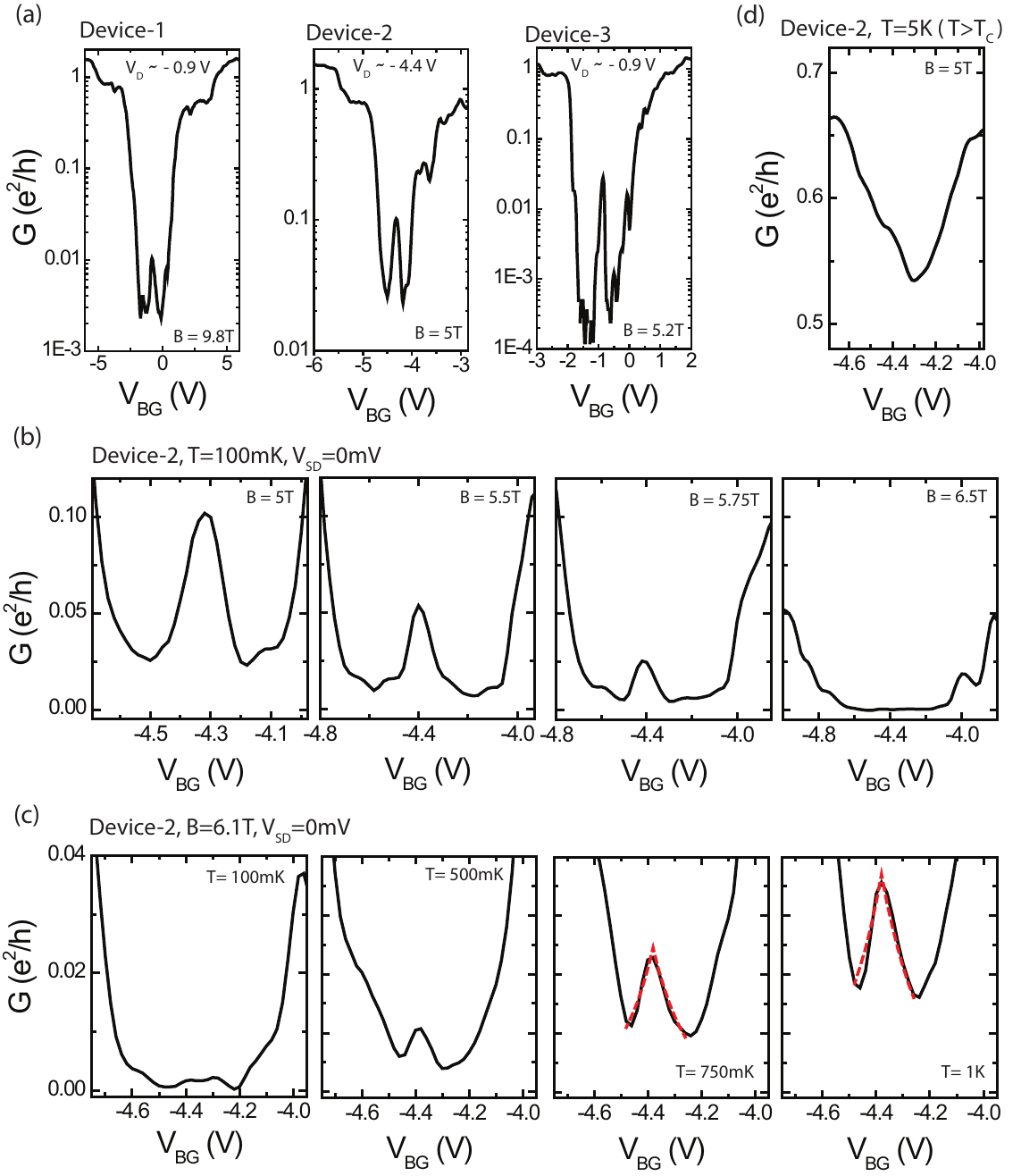}
 \caption{(Color Online) (\textbf{a}) The anomalous conductance peak at the DP shown in several devices on a log scale. (\textbf{b}) Conductance peak in device 2 at different magnetic fields shows the decrement of the amplitude with increasing B. (\textbf{c}) The conductance peak amplitude increases with increasing temperature. The red dashed lines in the last two panels display fitting of the peak line shape with Eq.~1. (\textbf{d}) No conductance peak at the DP is seen for T > $T_C$.}
 \label{fig:example}
\end{figure}

We begin by demonstrating that superconductivity in NbSe$_2$ survives up to high perpendicular magnetic fields where the uncovered graphene is comfortably in the QH regime. Fig. 1f shows the differential conductance (dI/dV) as a function of V$_{\rm SD}$, called the Andreev curve, for the values of V$_{\rm BG}$ marked A and B in Fig. 1d on the $\nu$=2 plateau. The existence of superconductivity is evident from the BCS like conductance peaks at about $\pm$0.5 meV for device 1 at B = 9.8T. Similar features are observed for device 2 (SM Fig. SI4-5f and Sec. SI6\cite{supplement}). Bias spectroscopy (SM Sec. SI6\cite{supplement}) allows us to extract the low-T superconducting gap ($2\Delta$) as a function of magnetic field, which we show in Fig. 4a; the large error bars arise primarily due to the asymmetric nature of the Andreev curve (the possible origin of which is discussed below). The superconducting gap of NbSe$_2$ flake, 2$\Delta \sim$ 2meV and T$_{C}$ $\sim$ 7K at 0T was directly characterized in our previous work (Fig. 3a of ref\cite{sahu2016andreev}), which is consistent with the 0T data in Fig. 4a. 
Fig.~1g shows the temperature dependence of the Andreev curves at B=9.8 T, which produces a T$_c \sim$ 2K where the BCS peaks disappear. We can relate the T$_c$ to 
superconducting gap through $2\Delta=4.07 k_BT_c \sim 0.7$meV (the factor 4.07 was determined in Ref.~\cite{rodrigo2004stm} for NbSe$_{2}$), which is close to that extracted from the Andreev curve at B=10T as shown in Fig. 4a. These observations -- appearance of BCS peaks in the Andreev curve (Fig. 1f) in a QH plateau and excellent agreement with the T dependence predicted by the BCS theory (Fig.~1g) -- demonstrate the coexistence of QH effect and superconductivity. It is noted that for bulk NbSe$_2$, the critical magnetic field is H$_{c2}\sim $4-5T\cite{xi2015ising}, but surface superconductivity (H$_{c3}$) has been reported for up to B=7-8T\cite{d1996evidence}; the existence of superconductivity at the interface of SLG-NbSe$_2$ at high magnetic field is thus not unexpected.

 \begin{figure}[ht!]
 \includegraphics[width=0.4\textwidth]{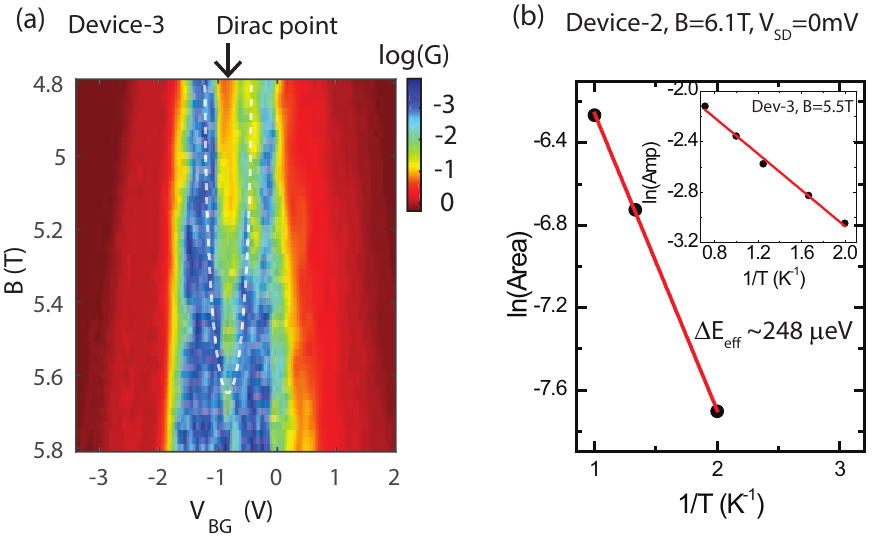}
 \caption{(Color Online) (\textbf{a}) 2D colormap of log(G) in device 3 plotted as a function of V$_{BG}$ and B showing the presence of the anomalous peak precisely at the DP, which vanishes above 5.6T. (\textbf{b}) Area of the peak plotted as a function of 1/T  showing  activated behavior with an effective gap of $\Delta E_{\rm eff}$ $\sim$ 248$\mu$eV. In the inset, amplitude of conductance peak in device-3 is used to show the activated behaviour, which gives $\Delta E_{\rm eff}$ $\sim$ 150$\mu$eV.}
 \label{fig:example}
\end{figure}

We next come to AR. Some evidence for it can be seen from the 
fact that the conductance at the 2e$^2$/h plateau is enhanced by 
$\sim$15\% (Fig.1h) when V$_{\rm SD}$ is changed from -3mV, where no AR is expected (because |eV$_{\rm SD}|>\Delta$), to zero, where AR is expected.
For an ideal, fully transparent contact, one expects 100\% enhancement due to AR; we attribute the smaller enhancement in our system to a non-fully transparent contact. Temperature dependence of conductance enhancement at $\nu$ = 2 is shown in SM Fig. SI4-5g\cite{supplement}. Conductance enhancement due to AR can also be seen by comparing the data below and above $T_C$ shown in SM Fig. SI4-5e\cite{supplement}. We note that the change in conductance for Andreev curve in Fig. 1f is around 10$\%$. However, the change of conductance was higher $\sim$ 25-30$\%$ for device2 in the QH regime (at $\nu$ = 2 plateau) as shown in SM Fig. SI6-8\cite{supplement}. At 0T the changes in Andreev curve was around 20$\%$ in device1 (SM Fig. SI6-7\cite{supplement}) and 45-50$\%$ in device2 (SM Fig. SI6-8\cite{supplement}).

Our most important finding is shown in Fig.~2, where a closer inspection of the conductance minimum reveals, completely unexpectedly, an anomalous peak. Further investigation brings out the following properties. First, the peak is seen precisely at the DP. Second, the peak is not seen above $T_C$ (compare Figs. 2d and 2c). Third, its amplitude decreases with decreasing temperature as well as increasing $\Delta E_I$, indicating that the peak is a finite temperature effect. Fig. 3a shows the 2D colormap of log(G) plotted as a function of V$_{\rm BG}$ and B, which displays the appearance of the peak precisely at the DP and its continuous decrement with increasing B. Finally, the parameters for which the anomalous peak is observed in device 2 and device 3 are shown by the dashed enclosed areas in the phase diagram in Fig. 4a; for both the devices the highlighted regime where the peak is observed satisfies the condition, $\Delta E_I<2\Delta$.

\begin{figure}[ht!]
\includegraphics[width=0.5\textwidth]{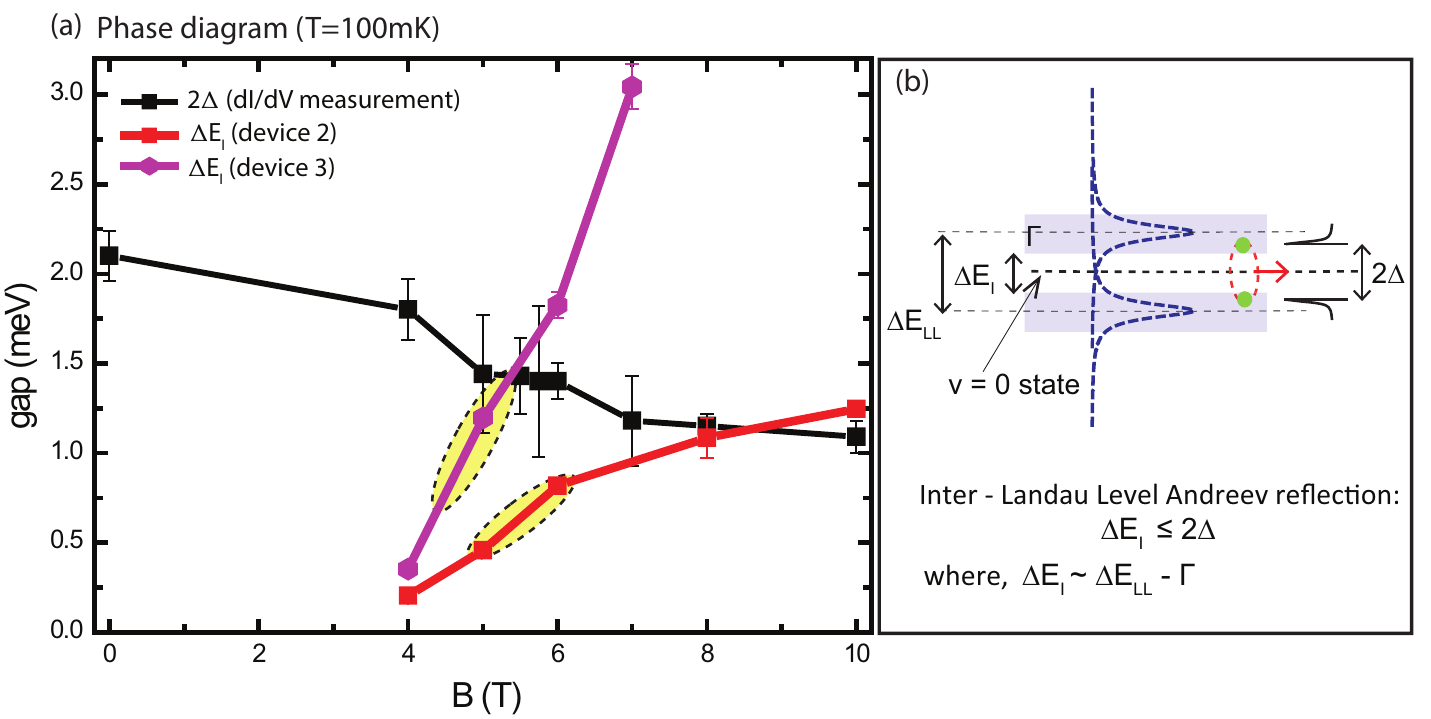}
 \caption{(Color Online) (\textbf{a}) An experimental phase diagram in energy and magnetic field. Filled black squares are the superconducting gaps measured using bias spectroscopy as a function of B. The filled red squares and filled purple hexagons show the insulating gaps of device-2 and device-3 as a function of B, where the thick lines are the guide to the eye. The anomalous conductance peak at the DP is observed in the region enclosed by the dashed black ovals. (\textbf{b}) Schematic of inter-Landau level AR process at the DP.}
 \label{fig:example}
\end{figure}

All of these facts are naturally explained in terms of a conductance peak originating from a new mechanism, namely finite temperature inter-Landau level AR, in which a thermally excited electron in the $N=0$ LL band above the E$_F$ 
reflects as a hole in the $N=0$ LL band below the E$_F$, as shown schematically in Fig. 4b. Such a peak is expected to occur (i) precisely at the DP, (ii) at finite temperature but for T$<$T$_c$, and (iii) for $2\Delta \geq \Delta E_I$. We mention that V$_{BG}$ at the DP depends slightly on whether the sweep is up or down, causing two different values in Fig. 2b; in Fig. 3a, all data are for sweep in the up direction, and show that the peak position remains invariant. We also note the presence of certain secondary, sample-specific peaks away from the DP, but their amplitudes are smaller by two to three orders of magnitude.

To see the activated nature of anomalous peak we plot the area under the peak in Fig. 3b for device 2, and fit it to a thermally activated behavior. Fitting the peak height gives a similar gap, as shown for device 3 in the inset of Fig. 3b. Further details regarding the activation nature of the peak for all the devices are shown in SM Sec. SI8 and SI9\cite{supplement}. Fitting the area in Fig. 3b using $e^{-\Delta E_{\rm eff}/2k_BT}$ gives $\Delta E_{\rm eff} \sim 0.25$ meV. One may expect $\Delta E_{\rm eff}$ to be equal to the $\Delta E_I$ (mobility gap), but the former is lower by a factor of $\sim 3$. This finds a natural explanation by the fact that the temperature dependence of the resistance of SLG shows two distinct $\Delta E_I$ differing by a factor of $\sim$ 3 (SM Sec. SI5\cite{supplement}): for example at B = 6T in device 2 for $T>2$K we have $\Delta E_I\sim 0.8$meV, but for $T<2$K we have $\Delta E_I \sim 0.25$meV, the latter being essentially in perfect agreement with the gap deduced from the anomalous peak at the DP. Similar results are obtained for device 3 as shown in SM Sec. SI5\cite{supplement}. Although the existence of the smaller, or `soft' gap around the $E_F$ in between the LLs at low temperature has been reported in the literature\cite{koch1995variable,giesbers2007quantum,efros1975coulomb,bennaceur2012unveiling}, its origin is not well understood. We ascribe the 'soft gap' below 2K to disorder.

\begin{figure}[ht!]
\includegraphics[width=0.5\textwidth]{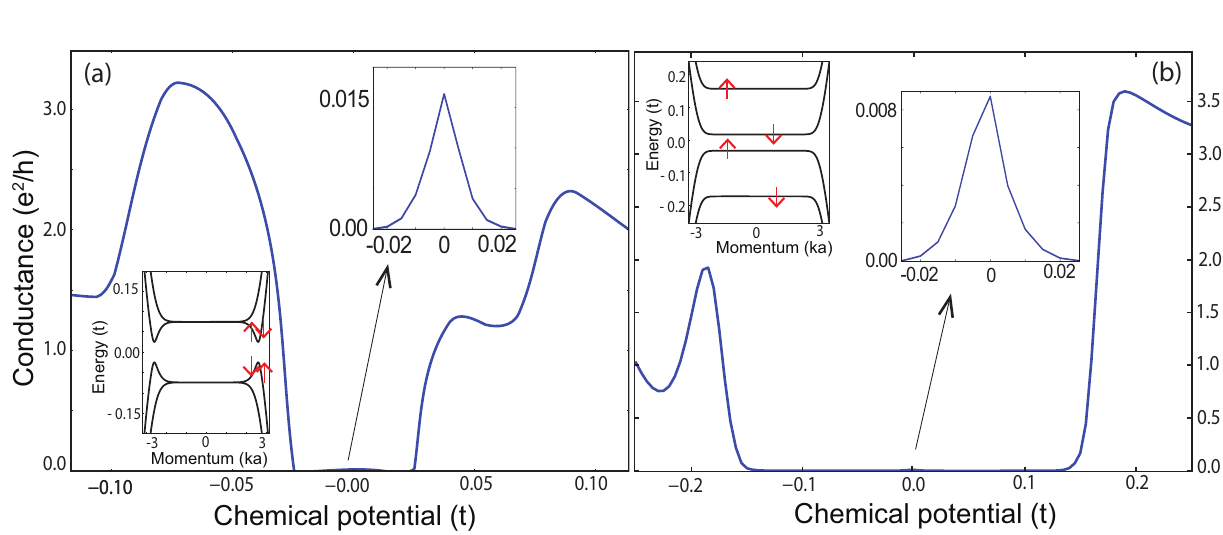}
 \caption{(Color Online) Panel \textbf{a} shows numerical results based on canted antiferromagnetic (CAF) model, and the panel \textbf{b} for the isospin ferromagnet (IFM) model. The  chemical potential is quoted in units of the hopping parameter $t$. The band diagram and the peak at the Dirac point are shown as insets. }
 \label{fig:example}
\end{figure}

To further confirm the physics of the inter-Landau level AR we have performed extensive numerical calculations, where we consider a system of graphene in the QH regime connected to superconducting graphene. The physics of the $\nu=0$ insulator at high B has been the subject of many studies\cite{zhang2006landau,checkelsky2008zero,du2009fractional,young2014tunable,bolotin2009observation,song2010high} and two most likely models are in terms of a canted antiferromagnet (CAF) or an isospin ferromagnet (IFM)\cite{kharitnov2012edge,abanin2006spin}, the band diagrams for which are schematically shown in the insets of Figs. 5a and 5b. The insulating gap of the former originates from a splitting of the $\nu=0$ LL into Landau bands with chiral edge states, whereas for the latter it results from a coupling between the helical edge states. To keep the discussion general, we consider AR in both models. The calculated conductance as a function of chemical potential (E$_F$) is plotted in Figs. 5a and 5b (SM-theory\cite{supplement} for the details) for CAF and IFM, respectively. It shows a small conductance peak at the DP arising from inter-Landau level AR (insets of Fig. 5a and 5b). At finite temperatures, the conductance at the DP can be analytically expressed as
\begin{equation}
G= \frac{e^2}{h}  \frac{2a}{1+e^{[\Delta E_I/2+|C(V_{BG}-V_D)|]/k_BT}} 
\end{equation}
where $a$ is the probability of AR and $C=dE_F/dV_{\rm BG}$. 
The experimental peak in Fig. 2c is fitted using the above equation with fitting parameters: a=0.35, $\Delta E_I$=0.5 meV, C=0.62 meV/V for T=1K and similar fitting is also shown for T = 0.75K. The fitting parameters are in general agreement with the experimental values (SM-theory\cite{supplement}).

Before ending, a comment on the physical origin of the observed asymmetry of the Andreev curves (Fig. 1f and SM Sec. SI6\cite{supplement}) is in order.
$dI/dV$ depends on the joint density of states (DOS) of the two materials. Typically, a normal metal has large and essentially constant DOS whereas the quasiparticle DOS of the superconductor is symmetric around zero bias, producing a symmetric Andreev curve. The density of states in a QH edge, in contrast, is complicated in real materials and can be energy dependent, thus producing asymmetric Andreev curves\cite{efetov2015specular,komatsu2012superconducting,park2017propagation,takayanagi1998semiconductor}.  We also note that due to the presence of the superconductor, the skipping orbits at the interface alternate between electron and hole-type orbits, whose centers are in general slightly offset (Fig. 1a bottom)\cite{hoppe2000andreev,PhysRevB.72.054518}, which  results in an interference pattern.
The fingerprints of the interference pattern can be seen as quasiperiodic conductance oscillations on the QH plateau as a function of the chemical potential (Fig. 1h and SM Sec. SI10\cite{supplement}). 
We refer the reader to previous literatures\cite{efetov2015specular,komatsu2012superconducting,park2017propagation,takayanagi1998semiconductor,morpurgo1999gate,finck2013anomalous,hoppe2000andreev,PhysRevB.72.054518} and the SM\cite{supplement} for details.

In conclusion, our primary accomplishment is an unambiguous demonstration of AR in graphene quantum Hall effect, which manifests most dramatically through an anomalous finite-temperature conductance peak at the Dirac point. By a combination of experimental and theoretical studies, we have  confirmed its origin as thermally induced inter-Landau level AR.

We thank Subhro Bhattacharjee, Tanmoy Das, H. R. Krishnamurthy, Subroto Mukerjee, Sumathi Rao, Sambuddha Sanyal, Ruchi Sexena, Vijay Shenoy, and Abhiram Soori for useful discussions. The authors acknowledge device fabrication and characterization facilities in CeNSE, IISc, Bangalore. A. D. thanks the Department of Science and Technology (DST), Government of India, under Grants No. DSTO1470 and No. DSTO1597. We also acknowledge the support by the U.S. Department of Energy, Office of Basic Energy Sciences, under Grant No. DE-SC0005042 (J. K. J.) and the National Key R\&D Program of China (Grant No. 2016YFA0401003) and NSFC [Grant No. 11674114 (X. L.)].

\bibliography{references}{}

\onecolumngrid
\newpage
\thispagestyle{empty}
\mbox{}


\section*{Supplementary material}

# SUPPLEMENTARY INFORMATION

The Supplementary Information consists of two sections, an experimental and a theoretical. In the former, we provide details of device fabrication and characterization, and also additional data for the determination of the insulating gap and the superconducting $T_c$. The theory section describes the model and gives calculational details.

## EXPERIMENTAL SECTION

### SI1 - Device fabrication and characterization

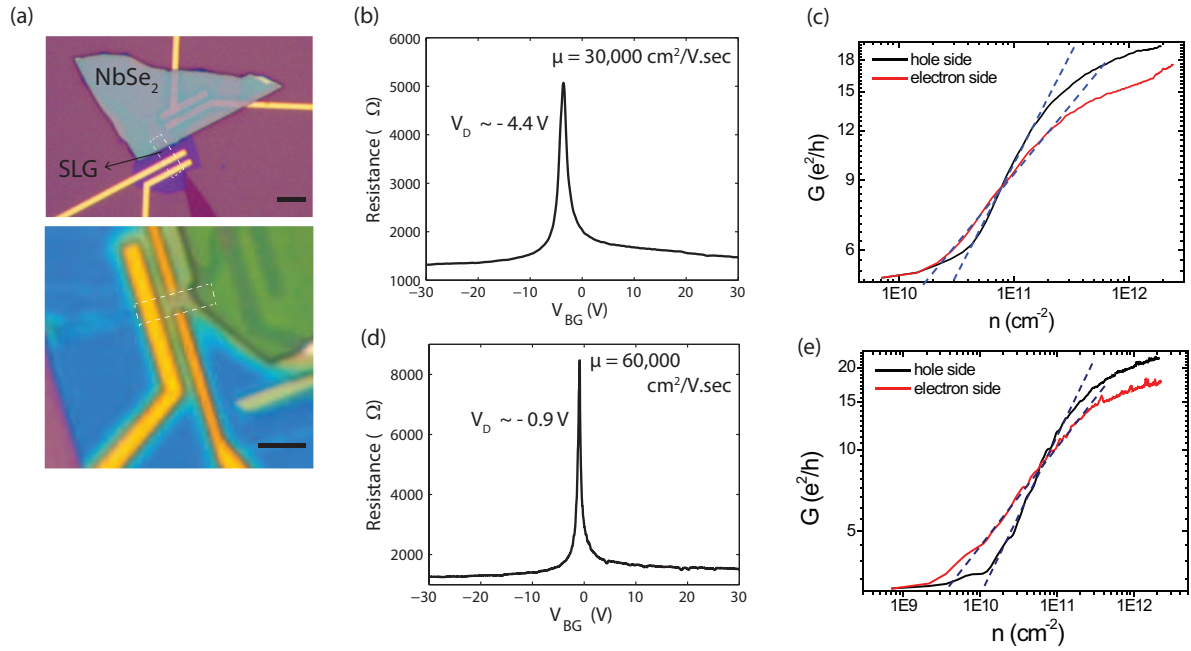

Figure SI1- 1: (**a**) Optical images of device-1 and device-2 with the scale bars of length $3\mu$m (**b**) Resistance in device-2 (Au - SLG - Au) at T=100mK as a function of the gate voltage. (**c**) Log-Log plot of conductance versus carrier density showing carrier in-homogeneity, $\delta n \sim (1-2) \times 10^{10} cm^{-2}$ (**d**) Resistance in device-3 (Au - SLG - Au) at T=100mK as a function of the gate voltage. (**e**) Log-Log plot of conductance versus carrier density showing carrier in-homogeneity, $\delta n \sim (3-5) \times 10^{9} cm^{-2}$ for electron side.

**Device fabrication:** Each device is fabricated by first exfoliating a thin hBN flake on an Si/SiO$_2$ substrate, followed by transferring an SLG on top of the hBN using polydimethylsiloxane (PDMS) based dry transfer

1

technique[1]. The contacts are made of Cr/Au(5nm/70nm) using the standard electron beam lithography technique followed by thermal deposition. Because NbSe$_2$ oxidizes when exposed to atmosphere, predefined contacts are made for NbSe$_2$, and at the final stage, the exfoliated NbSe$_2$ is transferred within a few minutes. Device-2 and device 3 were top-hBN protected with another stage of transfer to achieve higher mobility. Highest mobility of 60,000 cm$^2$/V.sec is achieved in device 3, where the carrier inhomogeneity ($\delta n$) is $\sim$ 3-5 $\times$ 10$^9$ cm$^{-2}$ for electron side, which gives a Fermi energy broadening $\delta E_F \sim$ 6-8 meV.

**Measurement technique:** Measurements were carried out in He3 cryostat as well as in a dilution refrigerator, with base temperatures of 240mK and 100mK, respectively. Standard lock-in technique is employed. All the measurements were performed using a low voltage bias of 20$\mu$V when measured in He3 cryostat, and 4$\mu$V when measured in dilution refrigerator.

**Measurement scheme:** The different measurement schemes used in our experiment are shown in Fig. SI1-2. For the R$_{XY}$ measurement, Current was injected between contacts A-D, whereas voltage was measured between contacts B-C. For the activation plots, the resistance was measured in two probe configuration, where voltage was applied at contact A and the current was measured at contact C. The R$_{XX}$ (to extract the LL broadening) was measured by injecting current between contacts B-D and measuring voltage between contacts A-C. For the Andreev reflection related measurements, the voltage was applied at contact A and current was measured at contact D.

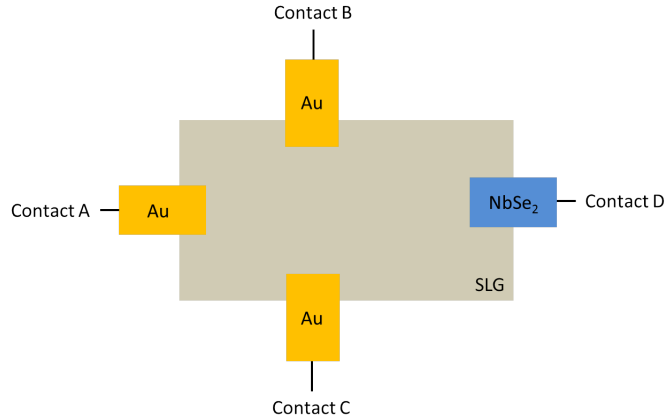

Figure SI1- 2: Measurement scheme

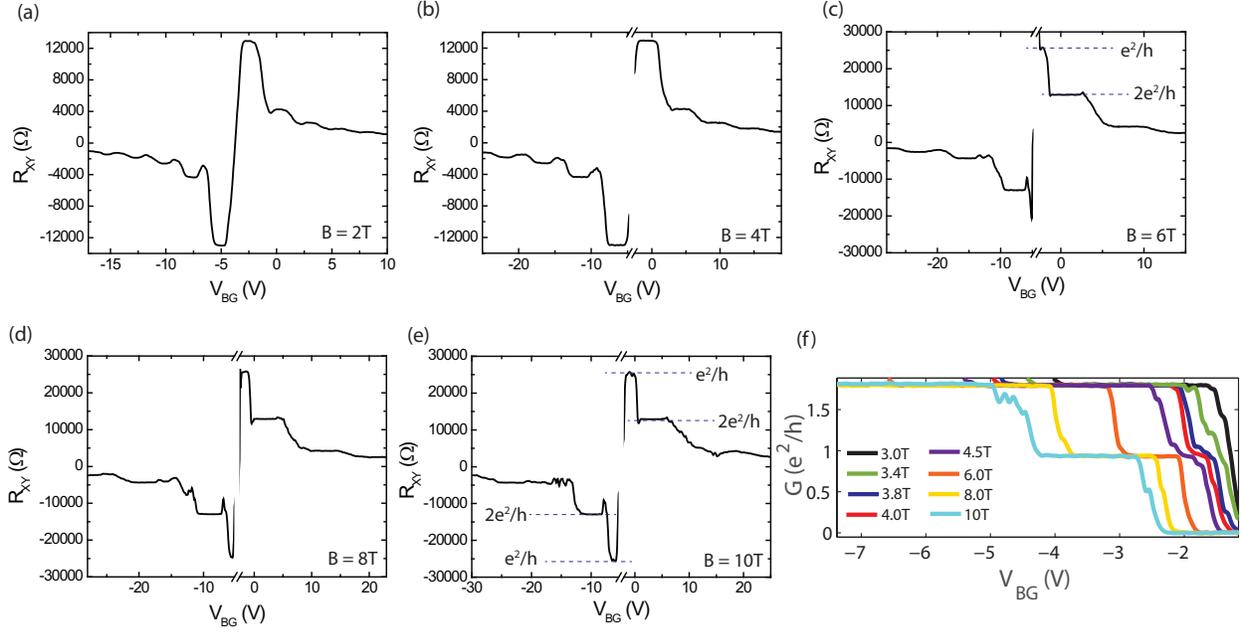

Figure SI2- 3: (**a-e**) $R_{XY}$ as a function of $V_{BG}$ at different magnetic fields measured in device-2 at T=100mK showing quantized resistance plateaus. (**f**) Two probe conductance of Au - SLG - Au in device 3 as a function of $V_{BG}$ for hole side only to show the $1e^2/h$ plateau clearly, the $1e^2/h$ plateau is observed above 3.8T indicating the high quality of the device.

### SI2 - QH response of device 2 and device 3

In Fig. SI2 we show the Hall resistance $R_{XY}$ in device-2. Well established quantum Hall plateaus are visible at B=2T indicating high device quality. Clear $1e^2/h$ plateau is visible at 10T; this plateau is identifiable in Hall measurement for B greater than 6T. For device 3 the $1e^2/h$ plateau is even visible at 3.8T as seen in Fig. SI2-f.

### SI3 - extraction of LL broadening

We have also evaluated the LL broadening ($\Gamma$) in device 3 from the magnetic field dependence of the amplitude of Shubnikov de Haas oscillations as described in Ref[2,3]. The average value of $\Gamma$ was found to be $\sim$ 4.5 meV, which is comparable to that in the device in Ref[3].

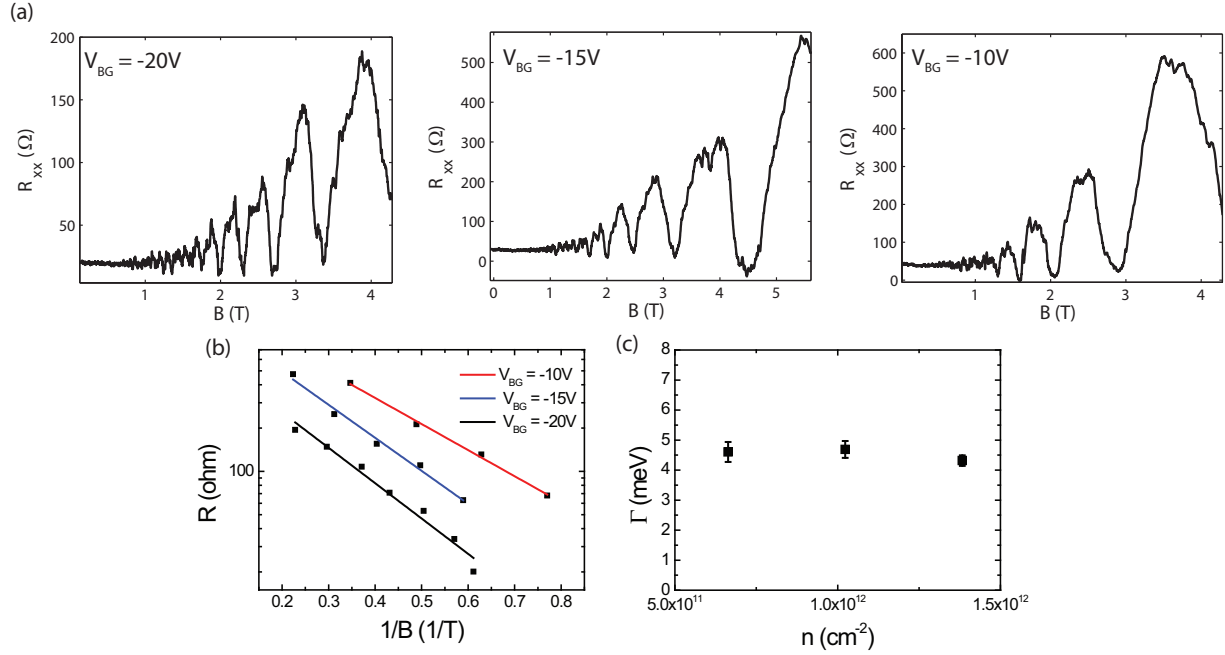

Figure SI3- 4: (**a**) Longitudinal resistance of device-3 plotted as a function of B showing conventional Shubnikov de Haas (SdH) oscillations. (**b**) Oscillation amplitude as a function of 1/B, slope is extracted from the linear fit. (**c**) Evaluated LL broadening ($\Gamma$) at different carrier concentrations.

### SI4 - QH response of the devices measured at Au-SLG-NbSe$_2$ contacts

Fig SI4(a-e) displays the two probe conductance of the SLG-NbSe$_2$ junction as a function of the backgate voltage in device-1 at T=240mK for several values of the magnetic field. In this device the 1e$^2$/h plateau appears at B=6T.

4

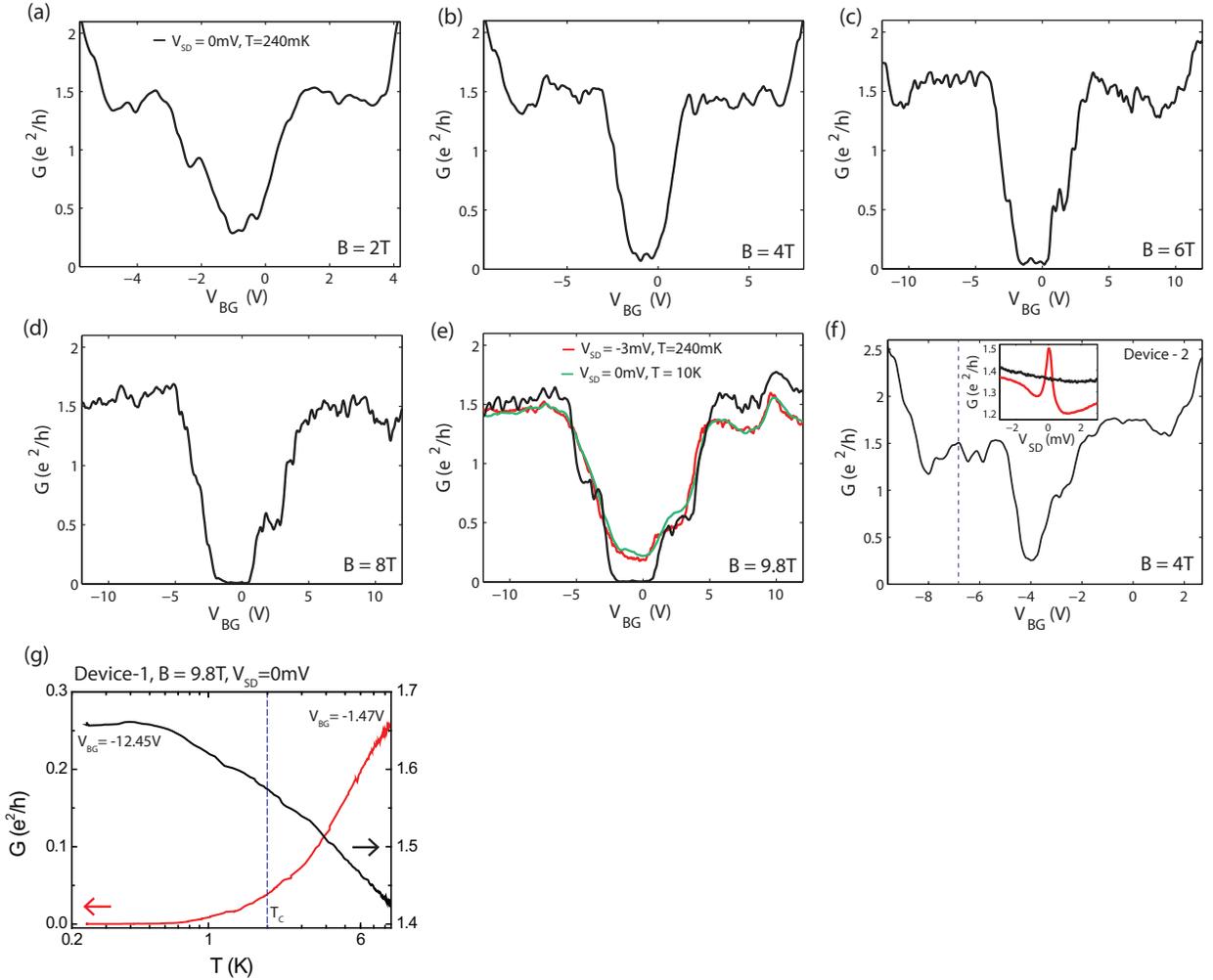

Figure SI4- 5: ((**a-e**) Two probe conductance of SLG-NbSe$_2$ junction in device-1 (Au-SLG-NbSe$_2$) as a function of the backgate voltage at several values of magnetic field, showing "quantized" conductance plateaus. Panel (e) has a comparison of the conductances at different bias voltages as well as different temperatures. (**f**) Two probe conductance of SLG-NbSe$_2$ junction in device-2 at B=4T showing "quantized" conductance plateaus. The inset shows the G-vs.-V$_{SD}$ plot at the V$_{BG}$ value marked by vertical dashed black line. The red line shows the conductance at T = 100mK and black line at T = 5K. The zero bias enhancement of conductance is correlated with the onset of superconductivity. (**g**) The conductance enhancement is quantified as a function of temperature at $\nu$ = 2 plateau (black curve). In contrast, the suppression of conductance at $\nu$=0 plateau is shown as a function of temperature (red curve).

5

## SI5 - Insulating gap at $\nu$ = 0

We have performed two probe measurement in device-2, in Au-SLG-Au configuration, to further understand the effects of broken valley and spin symmetries in graphene. In Fig SI5(a) the two probe gate response at T=100mK shows $\nu$=0 and $\nu$=1e$^2$/h plateaus at several B. SI5-b summarizes the insulating gap at $\nu$=0 plateau for device 2. SI5-c show the activation plots for device 1 at B=9.8T with insulating gas of $\sim$ 1 meV. The activation plot of device 3 at several magnetic fields is shown in SI5-d and e, where insulating gas of $\sim$ 5 meV is observed at B=10T. As mentioned in the manuscript the insulating gap (mobility gap) depends on the quality of the device, particularly on LL broadening. SI5-f and g show two distinct insulating gaps for T > 2K and T < 2K in device 2 and device 3, respectively. Although the smaller soft gap around the $E_F$ between the LL at low temperature has been known in the literature[4–7], exact origin for it in graphene is not clearly known. We ascribe the soft gap below 2K to disorder.

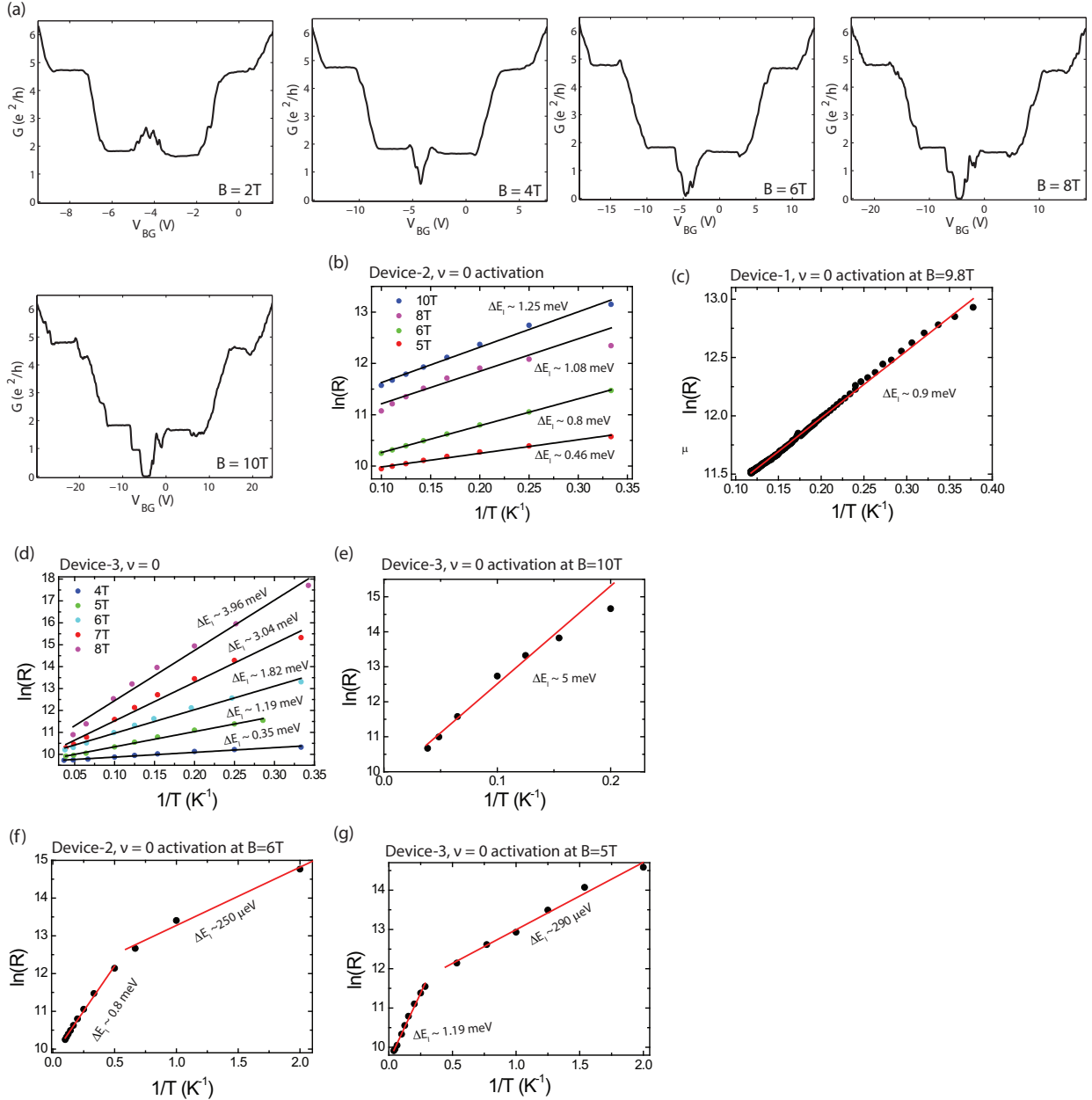

Figure SI5- 6: (a) Two probe conductance of SLG in device-2 (Au-SLG-Au) as a function of the backgate voltage at different magnetic fields, showing quantized conductance plateaus. A clear $\nu$=0 plateau is observed at B=8 and 10T. (b) Arrhenius plot in device 2 showing $\nu$=0 insulating gap at different magnetic fields. (c) Arrhenius plot in device 1 at B=9.8T. (d-e) Arhenius plot in device 3 at several magnetic fields. (f) Arrhenius plot in device 2 at B=6T showing two slopes corresponding to $\Delta E_I \sim$ 800 and 250 $\mu$eV for above and below 2K, respectively. (g) Arrhenius plot in device 3 at B=5T showing two slopes corresponding to $\Delta E_I \sim$ 1.19 meV and 290 $\mu$eV for above and below 2K, respectively.

7

## SI6 - BCS like features at different B

**Andreev curve at B=0:**

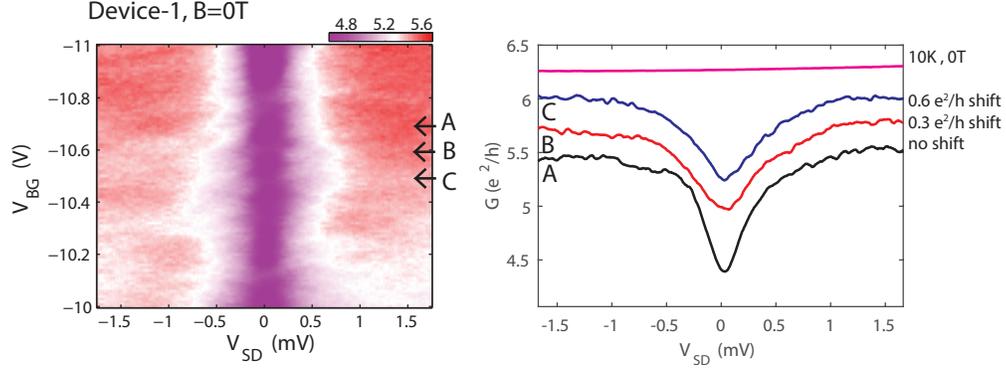

Figure SI6- 7: zero magnetic field diffrential conductance: (left)colormap of G plotted as functionn of $V_{BG}$ and $V_{SD}$ measured in device-1 at T=240mK at zero magnetic field. (right) G vs $V_{SD}$ at different gate voltages showing a conductance dip appearing due to Andreev reflection which is not present at 10K. The 10K curve differential conductance data (magenta) is at $V_{BG} = -1.6V$ and shifted by $4.2e^2/h$.

To evaluate the superconducting gap we have performed G versus $V_{SD}$ (Andreev curve) measurement, i.e., bias spectroscopy. $G(V_{SD})$ shows monotonic behavior for $T > T_C$, but begins to show non-monotonic features inside the superconducting gap below $T_C$. For ideal contacts with high transparency, theory of Andreev reflection predicts that the conductance should double within the superconducting gap, but in practice it can give a smaller enhancement depending on contact transparency. Fig. SI6-7 shows the differential conductance for device-1 at zero magnetic field; the conductance dip located precisely near zero bias, which is an usual signature of Andreev reflection, is not present at 10K.

Though the Andreev curve in zero magnetic field (Figs. SI6-7, SI6-8) is highly symmetric, it becomes asymmetric at finite magnetic fields. We discuss the possible sources of asymmetry below.

**Andreev curve at higher B:**

At high magnetic fields, the $\nu$=2 QH plateau is ideal for detecting the superconducting gap. The reason is that on a QH plateau, the QH edge states are ideally dissipationless and all of the voltage drop occurs at the

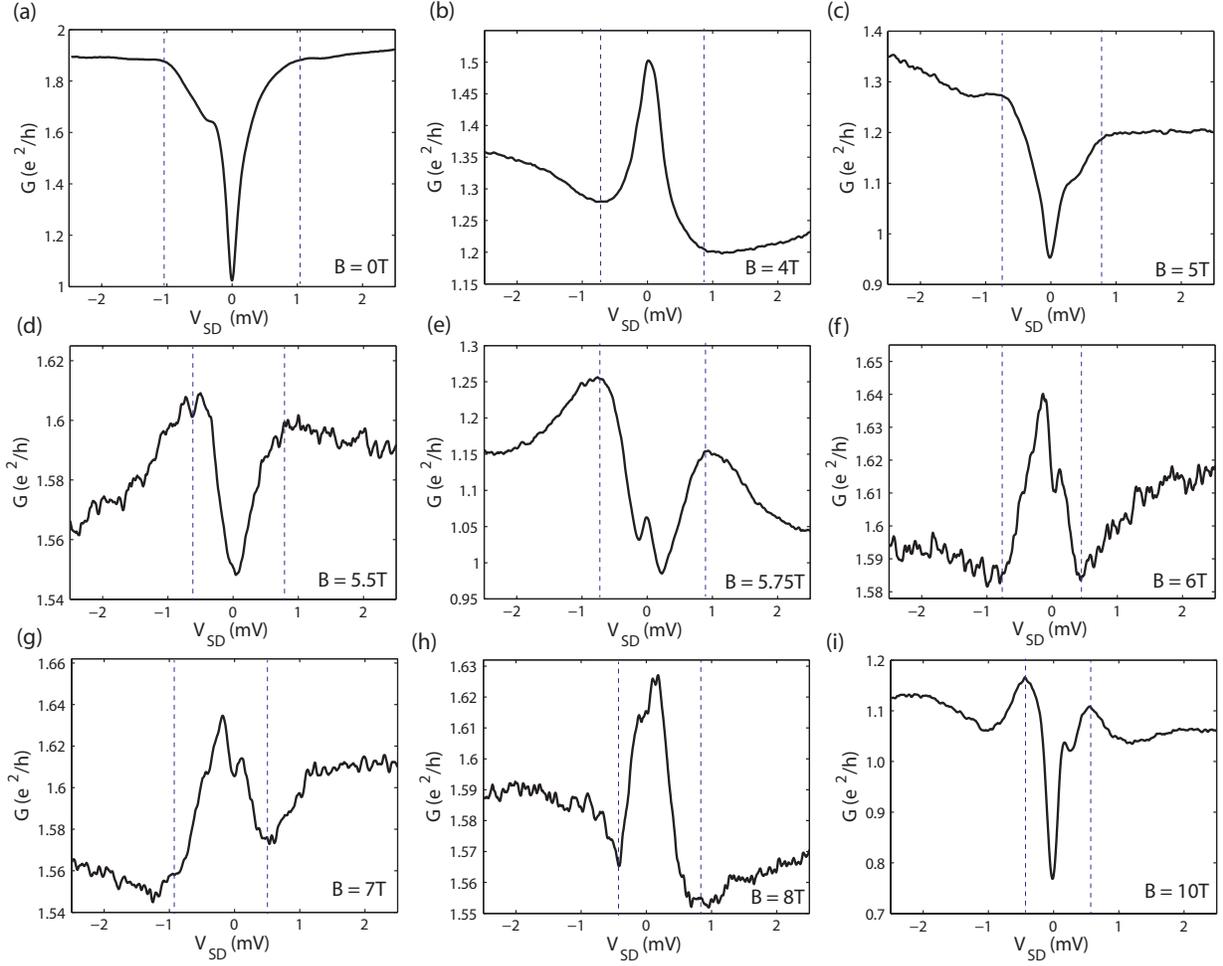

Figure SI6- 8: G vs $V_{SD}$ measurement in device-2 at T=100mK at different magnetic fields. BCS like features are evident. The separation between the conductance peaks/dips marked by vertical dotted lines yields the BCS gap (2$\Delta$) of the superconductor.

interface of SLG and SC. In both the devices, the G versus $V_{SD}$ plot at high magnetic fields produces either a zero-bias peak with dips on either side or a zero-bias dip with peaks on either side. The distance between the peaks or the dips yields the superconducting gap. It can be seen from Fig. SI5 that the superconducting gap decreases with increasing magnetic field, as expected.

**Origin of asymmetry in the high B Andreev curve:**

The differential conductance across a junction depends on the joint density of states (DOS) of the two

9

materials. In the case of a normal metal-superconductor junction, the normal metal has a large and essentially constant DOS, whereas the quasiparticle density of states in a superconductor is symmetric around zero bias. A convolution of these two results in a symmetric Andreev curve. In the presence of a magnetic field, when the chemical potential is in a QH plateau (between the Landau level), the density of states corresponding to the edge channels is quite complicated in a realistic sample, and can be energy dependent, leading to an asymmetric Andreev curve. Below $T_c$, another physics becomes relevant, namely the physics of conductance oscillations, predicted theoretically[8,9] and observed experimentally[10,11], which can further contribute to asymmetry of the Andreev curves. The underlying physics is that these oscillations depend on the wave vector (and thus the energy) of the incident electron, and therefore are not symmetric in source-drain bias. The physics of proximity induced oscillations is discussed in more detail in section SI10.

## SI7 - $T_C$ determination

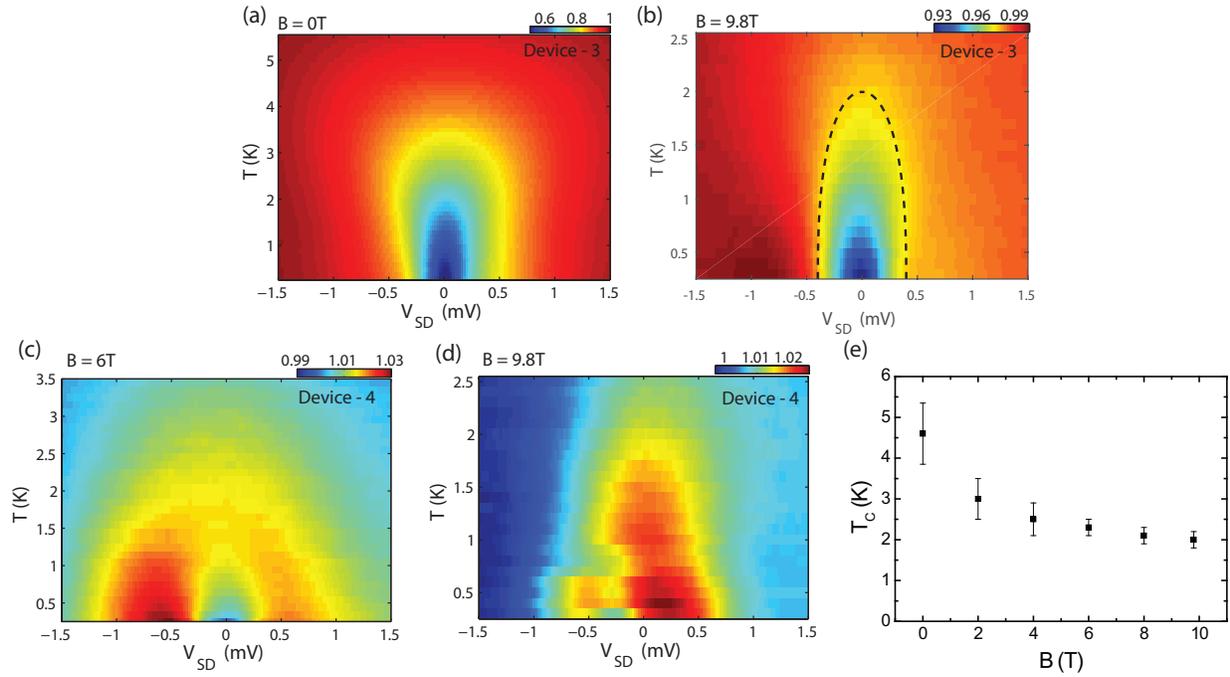

Figure SI7- 9: (a-d) 2D colormap of the normalised conductance (G($V_{SD}$)/G($V_{SD}$= -2 mV)) as a function of $V_{SD}$ and T at different magnetic fields. (e) $T_C$ as a function of magnetic field. In (b) the black dashed line shows the theoretical temperature dependence of superconducting gap calculated from BCS equation using parameters $2\Delta(T=0) = 0.8 meV$ and $T_C = 2K$.

As discussed in the previous section, the center of a plateau ($\nu$ = 2) is the best place to observe the effect of superconductivity in bias response. To evaluate the $T_C$ we have carried out bias measurements at

the centre of $\nu$=2 plateau at various temperatures. Fig SI7(a-d) shows the 2D colormap of the normalized conductance as a function of $V_{SD}$ and T at different values of B. The colormaps show vanishing superconductivity above a critical temperature. Similar measurement at other magnetic fields produces $T_C$ as a function of magnetic field, shown in fig SI7(e).

**SI8 - Activated nature of the peak at Dirac point**

In Fig SI8(a) the anomalous peak at the Dirac point, observed at B=10T in device-1, is shown at several temperatures. We have fitted the experimental data to a Lorentzian to extract the area under the peak, shown in Fig. SI8(b) as a function of T. The error bars indicate the quality of the fit. From the Arrhenius plot shown in Fig SI8(c), the log of area is seen to depend linearly on 1/T, which is a signature of an activated nature of the underlying process. The activation gap is estimated to be $\sim 180\mu$eV. Fig. 3a of the manuscript shows the phase diagram for device 2 and device 3, indicating the parameter range for the observation of the anomalous peak at the Dirac point. In Device-1 we could observe the peak at high magnetic field of 9.8T due to the fact that the mobility gap ($\sim 0.9$meV) and superconducting gap ($\sim 1$meV) are comparable at B = 9.8T. SI8-d shows activation nature of device 2, which has been plotted using peak height rather than area and shows an activation gap of $\sim 185\mu$eV, close to the value fitted with area as mentioned in the main text. SI8(e) shows the anomalous Dirac point peaks at several temperatures for device 3. In SI8(f) the area under the peak (device 2) is plotted as a function of $\Delta E_I$, which also shows activated nature.

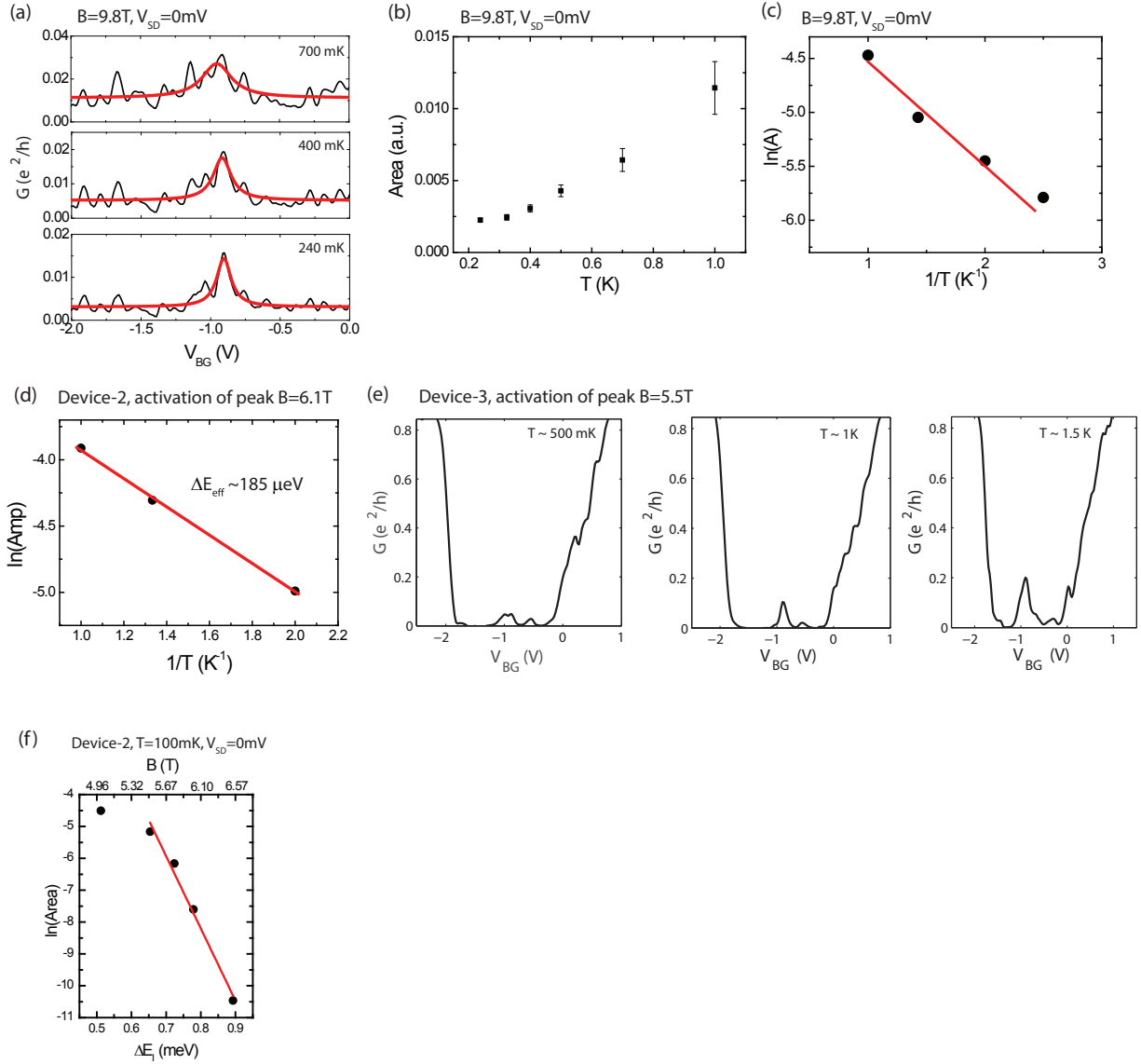

Figure SI8- 10: (**a**) G vs $V_{BG}$ plot fitted with a Lorentzian at different temperatures. (**b**) The area of the peak is plotted as a function of temperature. (**c**) Activated nature of the peak is evident in the log(area) vs. 1/T plot, which gives an effective gap of $\sim$ 180 $\mu$eV. (**d**) Activated behavior of amplitude of conductance peak in device 2. (**e**) Temperature evolution of conductance peak at B=5.5T in device 3. (**f**) Area of the peak at the DP is plotted as a function of insulating gap (which depends on B) showing activated behavior.

**SI9 - Evolution of the anomalous peak with magnetic field**

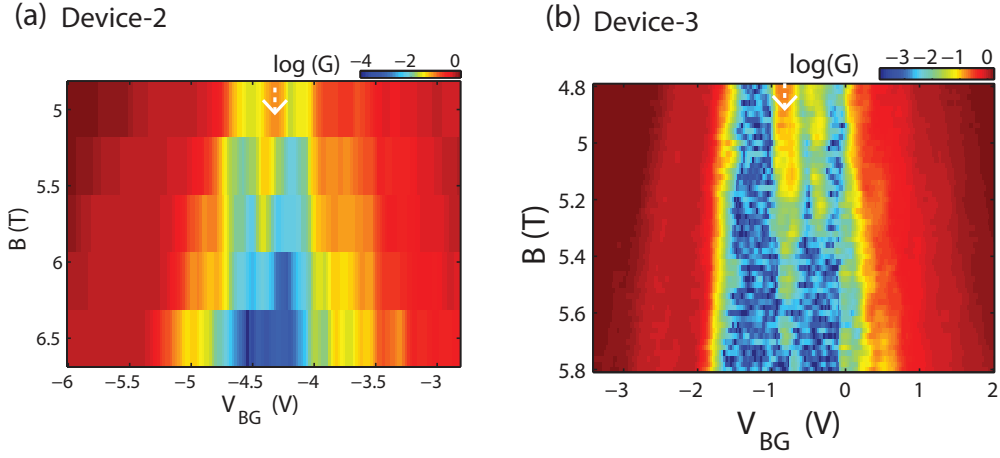

Figure SI9- 11: 2D colormap of log(G) plotted as a function of $V_{BG}$ and B in device 2 (a) and device 3 (b), respectively. The evolution of the anomalous Dirac peaks with the increasing insulating gap is clearly visible.

**SI10 - Conductance oscillations at $\nu = 2$**

The zero bias conductance on the $2e^2/h$ plateau exhibits (Fig. SI10) reproducible quasi-periodic oscillations of amplitude $\sim 0.2 e^2/h$ as a function of $V_{BG}$. Such oscillations are absent for $T > T_c$ (see SI4) as well as above superconducting gap, which strongly suggests that these are a manifestation of the Andreev physics. Along the junction interface (Fig. SI10(b)) the centers of the electron and the hole trajectories (classically, the radii of skipping cyclotron orbits) are offset by a distance $d$ of the order of the magnetic length $l_B = \sqrt{\hbar/eB}$, thereby defining an area $\sim d \times W$ ($W$ being the sample width), which can give rise to periodic oscillations as function of chemical potential($\mu$) and magnetic field(B)[8,9]. we have also observed similar conductance oscillations as a function of magnetic field (Fig. SI10(c)). Because of this Aharonov- Bohm like effect the Andreev reflection at the interface of graphene QH and superconductor is more intriguing and its effect is observed in our experiment in form of both conductance oscillations and peak or dip in differential conductance plot, which is consistent with the literatures[10,11]. We note, however, that the effect of disorder cannot be ruled out and might be responsible for the absence of nice periodic oscillations, as expected from the interference physics discussed above.

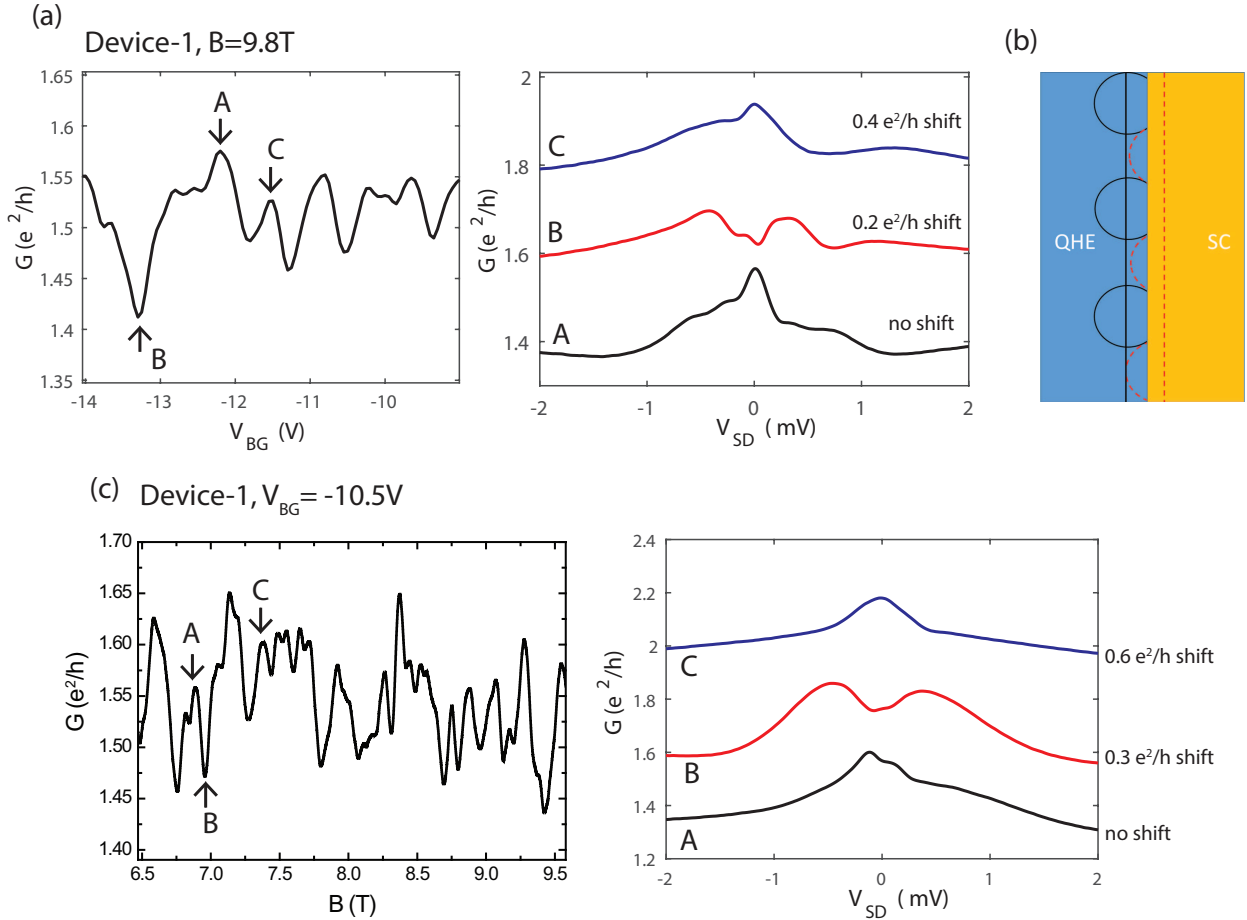

Figure SI10- 12: (a) conductance plotted as a function of $V_{BG}$ (left) showing quasi periodic oscillations on $\nu = 2$ QH plateau at B=9.8T and the differential conductance plotted as a function of $V_{SD}$ (right) at peaks or dips marked in the gate response curve. (b) Classical skipping orbits of electron and hole at the interface; vertical lines show the center of the orbits. (c) conductance plotted as a function of B (left) showing similar quasiperiodic oscillations on $\nu = 2$ QH plateau at $V_{BG} = -10.5V$ and the differential conductance plotted as a function of $V_{SD}$ (right) at peaks or dips marked in the G vs B curve.

## THEORETICAL SECTION

**SI11 - Andreev reflection in quantum Hall regime**

In the metal-superconductor junction, the electrons and holes on the metallic side are coherently coupled to those in the superconductor by Andreev reflection at the metal-superconductor interface, which is the origin of the superconducting proximity effect. Here, we theoretically demonstrate that such electron-hole coherence is maintained even when the metal is in the quantum Hall regime, where only the chiral edge states are available at the Fermi energy.

Our model is as follows. A planar interface is considered, located at $x = 0$ between a semi-infinite region $x > 0$ occupied by a NbSe2 superconductor and a semi-infinite graphene region ($x < 0$). A uniform magnetic field is applied in the z direction, which we assume is screened for $x > 0$ by the superconductor due to the Meissner effect. We thus assume an abrupt change of the magnetic-field strength at the interface: $B(x) = B(1 - \Theta(x))$, where $\Theta(x)$ is the step function. In the basis $\psi_i = (c_{i\uparrow}, c_{i\downarrow})^\mathrm{T}$, the Hamiltonian of the graphene in the presence of a magnetic field can be generally written as

$$
\begin{aligned}
H &= H_0 + H_I \\
H_0 &= \sum_i c_i^\dagger (-\mu \sigma_0 + M_z \sigma_z) c_i + \sum_{\langle ij \rangle} -t e^{-i\phi_{ij}} c_i^\dagger c_j \\
H_I &= \sum_i c_i^\dagger (\epsilon_i \sigma_0 + u_i \sigma_x) c_i.
\end{aligned}
\tag{1}
$$

Here $H_0$ is the Hamiltonian for perfect graphene in the presence of a transverse magnetic field. $M_z$ denotes the Zeeman splitting due to the applied external magnetic field. At the maximal magnetic field in our experiment (10T) we have $M_z \approx 1$meV, which is of the same order as the superconducting gap. The $H_0$ Hamiltonian does not capture the physics at $\nu = 0$, because it produces a gapless spin ferromagnet, rather than a gapped insulator as observed experimentally. We incorporate this physics by adding an anisotropic term Hamiltonian $H_I$, which is due to either the sublattice anisotropy ($\epsilon_{i=a} = -\epsilon_{i=b}$) [3] or the in plane anisotropy ($u_{i=a} = -u_{i=b}$) [12]. This term makes the system fully gapped at Dirac point. The sublattice anisotropy $\epsilon_i$ and $u_i$ lead to the so called isospin ferromagnet (IFM) and canted anti-ferromagnet (CAF) separately.

We first consider Andreev reflection near the $\nu = 0$ Landau level. Although the origin of the observed insulating gap in this regime is still being debated, we find that both the IFM and CAF models produce a sharp conductance peak at Dirac point. This peak serves as a smoking gun signature of Andreev reflection at QH-superconductor interface, because the system near $\nu = 0$ Landau levels can only have interband Andreev reflection which is independent of the specific form of the insulating gap. Let us now explain the basic physics.

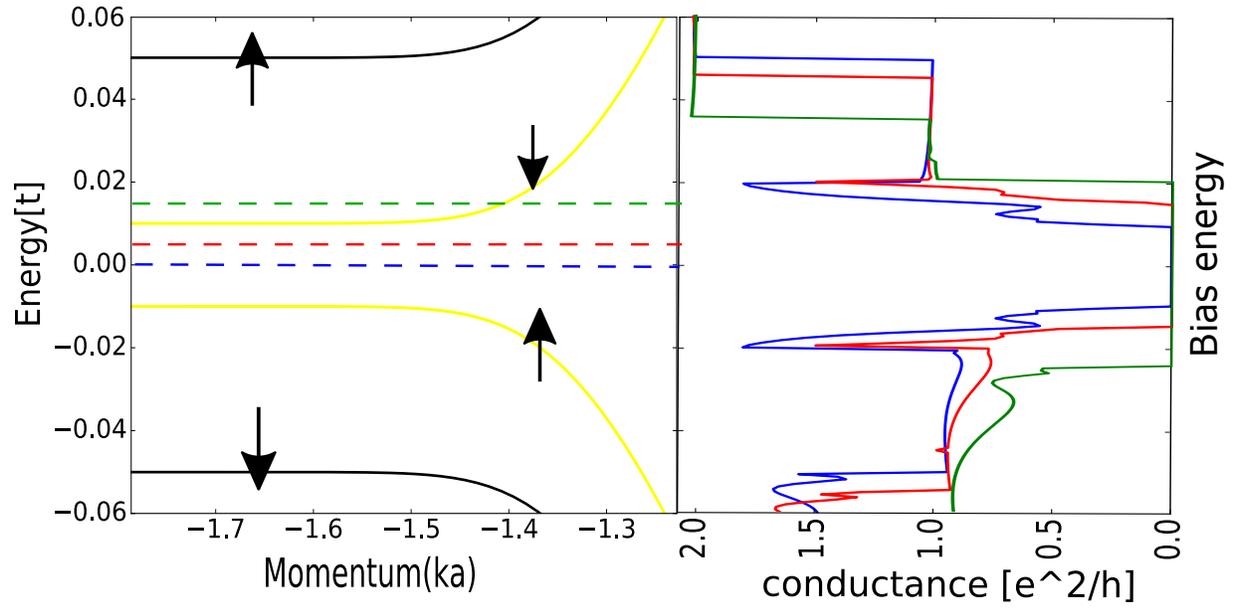

Figure SI11- 13: Left panel: Band dispersion for an isospin ferromagnet (IFM). The breaking of the sublattice symmetry through $\epsilon_{ia} = -\epsilon_{ib} = \epsilon$ (a and b indicating the two sublattice of the graphene nanoribbon) provides a band gap at 0th Landau level. Right panel: The differential conductance of the QHE-superconductor junction with different chemical potentials. Blue, red and Green curves corresponds to the three chemical potentials marked by horizontal blue, red and green dashed lines in the left panel.

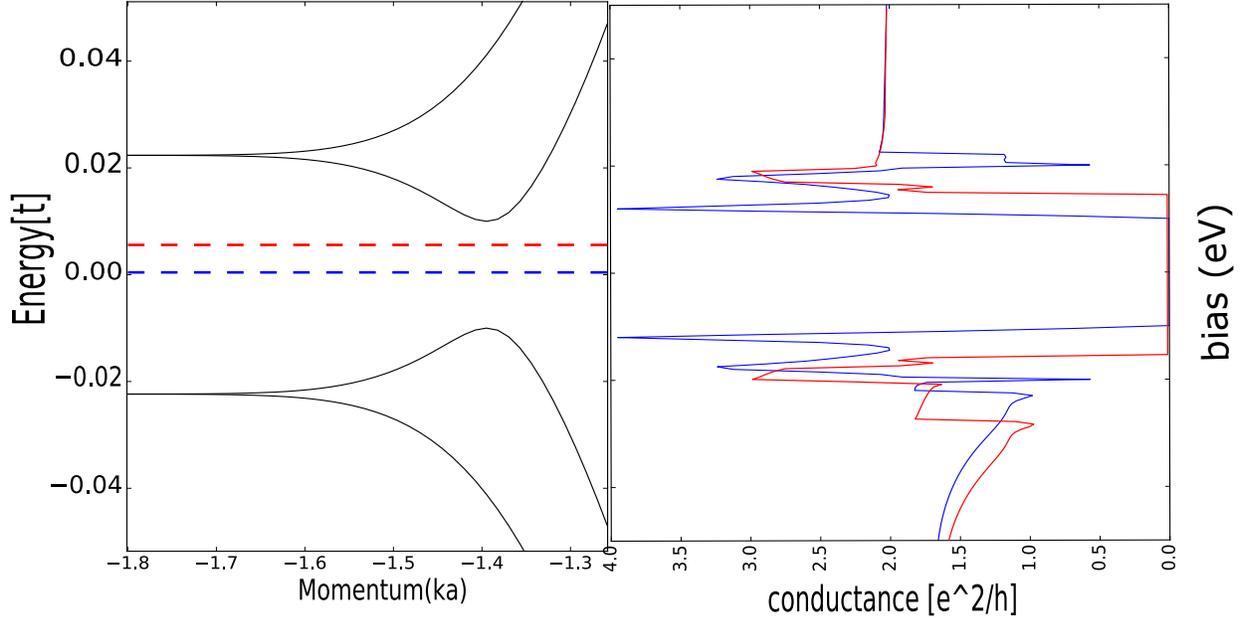

Figure SI11- 14: Left panel: Band dispersion for a canted antiferromagnet (CAF). The in-plane anisotropy term $u_{ia} = -u_{ib} = \epsilon$ (a and b indicating the two sublattice of the graphene nanoribbon) provides a band gap at 0th Landau level. Right panel: The differential conductance of the QHE-superconductor junction for two different chemical potentials, colour matched to the chemical potentials marked by the horizontal dashed lines in the left panel.

The IFM insulator has a band gap $2\Delta_I$ with $\Delta_I = (\epsilon - Mz)$. Let us first consider the situation at zero temperature. When the chemical potential of graphene is zero, the Andreev reflection can happen when the bias voltage is larger than the threshold value $\Delta_I/e$ which leads inter-band Andreev reflection between the bottom electron and top hole Landau levels (yellow bands in Fig. ). When the system has a finite chemical potential $\mu$ but still lies in the gap ($|\mu| < \Delta_I$, e.g. the red dashed line in Fig. SI11-13, the threshold voltage increases to $(\Delta_I + |\mu|)/e$, as the Andreev reflection couples two electrons with opposite energy. This indicates that the threshold bias voltage is minimum at chemical potential $\mu = 0$, which results, at nonzero temperatures, in a anomalous peak at precisely the Dirac point.

For a more quantitative account, we obtain the conductance at zero temperature from well-known Landauer-Büttiker formula for the SC/normal-metal junction, shown in the right panel of Fig. SI11-13. At finite temperatures, the current has the form

$$I = \frac{e}{h} \int_{-\infty}^{\infty} f(E)(1 - r_{ee}^2 + r_{he}^2) dE, \qquad (2)$$

where $f(E)$ is the Fermi-Dirac distribution function, $r_{ee}$ and $r_{he}$ are the amplitude of normal reflection and amplitude of Andreev reflection respectively. $r_{he}$ will be non-zero only beyond $E = E_F + (\Delta_I + |\mu|)$, using

17

this information for infinitesimal excitation voltage $dV$ current will take the form

$$I = \frac{e}{h} \frac{2a}{1 + e^{(\Delta_I + |\mu|)/kT}} edV, \tag{3}$$

where $a = r_{he}^2 = 1 - r_{ee}^2$ is the probability of Andreev reflection. Finally the conductance (G) will be given by

$$G = \frac{e^2}{h} \frac{2a}{1 + e^{(\Delta_I + |\mu|)/kT}}, \tag{4}$$

We have shown that the anomalous Dirac peak evaluated from this expression is well consistent with the experimental peak. It is important to note that the conversion of gate voltage to $\mu$ is not straightforward in the insulating regime and requires a detailed knowledge of density of states in the gap. Assuming linear dependence of chemical potential on $V_{BG}$ in the narrow region of our interest, we can rewrite $\mu = C(V_{BG} - V_D)$, where C is the change in chemical potential per unit change in $V_{BG}$ ($dE_F/dV_{BG}$). In our analysis we have used C as a fitting parameter, which yields the value $\sim 0.6$ meV/V. This number is of a reasonable magnitude because: in device1 the $\nu = 0$ plateau has a width of 2V in the gate voltage and from activation we have extracted a gap of $\sim 1meV$; in device 3 at 10T the $\nu = 0$ plateau width is $\sim 3$ Volts, which produces an insulating gap of $\sim 5meV$.

A CAF has a band gap $2\Delta_I$ with $\Delta_I = |u|$. In spite of the different microscopic mechanism for the origin of the insulating behaviour, the threshold voltage for interband Andreev reflection still increases as the chemical potential moves away from the Dirac point, as shown in Fig. SI11-14 from explicit calculation. This also leads to a differential conductance peak at the Dirac point.


\end{document}